\documentclass[journal]{IEEEtran}
\usepackage{amssymb}
\ifCLASSINFOpdf
\else
\fi
\usepackage{graphicx}

\usepackage{cite}

\usepackage{pifont}
\usepackage{bbding}
\usepackage{amsmath}
\usepackage{amssymb}
\usepackage{mathrsfs,balance}

\begin{document}
%
\title{Joint Optimization of Power Splitting and Allocation for SWIPT in Interference Alignment Networks}
%
%
%

\author{Nan~Zhao,~\IEEEmembership{Senior Member,~IEEE}
\thanks{N. Zhao is with the School of Information and Communication Engineering, Dalian University of Technology, China (email: zhaonan@dlut.edu.cn).}


}

\maketitle

\begin{abstract}
Interference alignment (IA) is a promising solution for interference management in wireless networks. On the other hand, simultaneous wireless information and power transfer (SWIPT) has become an emerging technique. Although some works have been done on IA and SWIPT, these two important areas have traditionally been addressed separately in the literature. In this paper, we propose to use a common framework to jointly study IA and SWIPT. We analyze the performance of SWIPT in IA networks. Specifically, we derive the upper bound of the power that can be harvested in IA networks. In addition, we show that, to improve the performance of wireless power transfer and information transmission, users should be dynamically selected as energy harvesting (EH) or information decoding (ID) terminals. Furthermore, we design two easy-implemented SWIPT-user selection (SWIPT-US) algorithms in IA networks. To optimize the ID and EH performance of SWIPT in IA networks, a power-splitting optimization (PSO) algorithm is proposed when power splitters are available, and its closed-form optimal solutions are derived. Power allocation in the PSO algorithm is also studied to further optimize the performance. Simulation results are presented to show the effectiveness of the proposed algorithms.

\end{abstract}

\begin{IEEEkeywords}
Simultaneous wireless information and power transfer, interference alignment, energy harvesting, information decoding, power splitting, power-to-rate ratio.
\end{IEEEkeywords}

%
\IEEEpeerreviewmaketitle

\section{Introduction}

\IEEEPARstart{I}{nterference} management is one of the key issues in wireless networks. Recently, \emph{interference alignment} (IA) has become a promising solution to the  interference problem in wireless networks \cite{Cadambe08, Maddah-Ali08,Zhao11}. With IA, the transmitted signals are designed to cooperatively constrain all the interferences into certain subspaces at the unintended receivers, and thus the  remaining interference-free subspaces can be exploited to obtain the desired signal at each receiver \cite{Jafar10}.

Some challenges of applying IA in practical networks were presented in \cite{Ayach2013}, and plenty of effort has been conducted to make IA more practical \cite{Gomadam11,Santamaria10, Guler11,TWC,analogCSI, Tian2013, Minn2013, Zhaosubmitted2012,CSI1,ASLi2016,NanPLS,7565184,6918468,zhao2016adaptive}. In \cite{Gomadam11}, iterative IA algorithms based on the reciprocity of wireless networks were designed, which make IA easier to realize.
The performance degradation in IA networks at low signal-to-noise ratio (SNR) was analyzed in \cite{TWC}, and several research works have been presented to improve the sum rate or QoS of IA networks \cite{Gomadam11, Santamaria10, Guler11, TWC}. Since accurate channel state information (CSI) should be available at the transmitters for IA, the authors of \cite{analogCSI, Tian2013, Minn2013, Zhaosubmitted2012,CSI1} focused on solving the problem of imperfect CSI in IA networks.


 In wireless networks, another key issue is \emph{energy} consumption. Due to the rapidly rising energy costs and global CO$_2$ emissions, green communications have attracted a lot of interest from both academia and industry \cite{greencommunications, xie2012}.
 Energy harvesting (EH) is an important method to achieve green communications \cite{TGC13}. In EH, the energy captured from external sources is converted to electrical energy to support green self-sufficient wireless nodes \cite{Ozel2011, Orhan2013, Zorzi14}.
 Since radio-frequency (RF) signals carry energy, RF can be a new source for energy harvesting. Indeed, wireless power transfer (WPT) through RF signals can be applied to the scenarios with low-power applications, and thus becomes an important aspect of energy harvesting \cite{Science07, Huang14}.
   As RF signals are commonly used as a vehicle for transmitting information in wireless networks, \emph{simultaneous wireless information and power transfer} (SWIPT) has become an emerging technique attracting great attention \cite{Varshney2008, Grover10, rzhang2013, RZhangTCOM2013, XT12, SimeoneTCOM13, HuangTSP13, FouladgarArxiv}.

 Some initial works in SWIPT have been conducted in \cite{Varshney2008, Grover10} to analyze the maximal information rate versus (vs.) energy performance in single-input single-output (SISO) channel without and with frequency-selective fading, respectively.
Zhang \emph{et al}. \cite{rzhang2013} studied information rate vs. energy performance in a multi-input multi-output (MIMO) wireless broadcast system consisting of one EH receiver and one information decoding (ID) receiver.
  In \cite{RZhangTCOM2013}, the optimal power splitting was performed in SISO and single-input multiple-output (SIMO) systems to achieve various trade-offs between information capacity and energy harvested. Xiang \emph{et al}. \cite{XT12}  studied the robust beamforming problem for three-node multiple-input single-output (MISO) broadcasting systems with one energy receiver and one information receiver when the CSI is imperfect.
 A two-way communication system with two nodes communicating in an interactive fashion to achieve SWIPT was studied in \cite{SimeoneTCOM13}, where information ``1" and ``0" correspond to one unit of energy and no energy transferred, respectively. In \cite{HuangTSP13}, a practical framework for SWIPT in broadband wireless systems was developed through exploiting orthogonal frequency division multiplexing (OFDM) systems to divide a broadband channel into decoupled narrowband sub-channels, where frequency diversity is also utilized  to improve the efficiency of SWIPT. Fouladgar \emph{et al}. \cite{FouladgarArxiv} proposed to use constrained run-length limited codes instead of conventional unconstrained codes in SWIPT to limit battery overflow and control battery underflow.

 Although some excellent works have been done on IA and SWIPT, these two important areas have traditionally been addressed separately in the literature \cite{IWCMC,NanWC2014,NanEH,SWIPT3}. For example, in IA networks, the interferences are usually leveraged to separate out the desired signal instead of reutilization, which is a great waste of energy in wireless networks. On the other hand, in the existing SWIPT studies, recent advances in IA are largely ignored.

In this paper, we propose to use a common framework to jointly study IA and SWIPT. The main contributions of this work can be summarized as follows.
 \begin{itemize}
\item SWIPT is becoming an important aspect in the research of EH, however, to the best of our knowledge, the SWIPT issue in IA networks is largely ignored in the existing works. Thus a unified framework of SWIPT in IA is introduced, and some fundamental work is presented in this paper.
\item We analyze the performance of SWIPT in IA networks. Specifically, we derive the lower and upper bounds of the power that can be harvested in IA networks.
\item To improve the performance of wireless power transfer and information transmission, we show that users should be dynamically selected as EH or ID terminals. Then, we design two easy-implemented SWIPT-user selection (SWIPT-US) algorithms in IA networks. The first one is based on the round-robin principle, which is natural and simple. To further improve the performance, we define a parameter called power-to-rate radio (PRR), and design a PRR-based SWIPT-US algorithm.
\item We propose a power-splitting optimization (PSO) algorithm to optimize the ID and EH performance of IA networks when power splitters are available, in which the specific rate and energy requirements of  users are taken into account. The closed-form optimal solutions to the PSO algorithm are derived. In addition, the power allocation (PA) problem in the PSO algorithm is studied.
\item Simulation results are presented to show the effectiveness of the proposed algorithms.

\end{itemize}

The rest of the paper is organized as follows. In Section II, the system model is presented. The performance of SWIPT in IA networks is analyzed in Section III. In Section IV, two SWIPT-US algorithms are proposed for SWIPT in IA networks. In Section V, the PSO algorithm is proposed to optimize the ID and EH performance of IA networks, and power allocation is also studied. Simulation results are discussed in Section VI, and finally, conclusions and future work are presented in Section VII.

\emph{Notation:} $\textbf{I}_\emph{d}$ represents the ${d}\times{d}$ identity matrix.
 $\textbf{A}^{\dagger}$  and $|\textbf{A}|$ are the Hermitian transpose and determinant of matrix $\textbf{A}$, respectively. $\|\textbf{a}\|_2$ and $\|\textbf{A}\|_2$ are the $\ell^2$-norm of vector $\textbf{a}$ and matrix $\textbf{A}$, respectively. ${{\lambda _{\max }}\left({\textbf{A}} \right)}$ means the maximal eigenvalue of the matrix $\textbf{A}$. $|a|$
is the absolute value of complex number \emph{a}. For two integers $b$ and $c$, $b\%c$ means $b$ modulo $c$. $\mathbb{C}^{M\times N}$ is the space of complex $M\times N$ matrices. $\mathcal{CN}(\textbf{a}, \textbf{A})$ is the complex Gaussian distribution with mean \textbf{a} and covariance matrix \textbf{A}. $\mathbb{E}(\cdot)$ stands for expectation.

\section{System Description\label{sect: system}}
In this section, we first present the model for linear IA wireless networks. Then, wireless power transfer in IA networks is introduced.
\subsection{Linear Interference Alignment Wireless Networks}
Consider a $K$-user interference channel with $M^{[k]}$ and $N^{[k]}$ antennas equipped at the $k$th transmitter and receiver, respectively. Perfect CSI of the network is assumed to be available at all the transceivers. If linear IA is adopted to avoid interferences among users, the received signal with $d^{[k]}$ data streams at receiver $k$ can be represented as
\begin{eqnarray}
\!\!\!\!\!\!\textbf{y}^{[k]}\!(n)\!\!\!\!\!&=&\!\!\!\!\!\textbf{U}^{[k]\dagger}(n)\textbf{H}^{[kk]}(n)\textbf{V}^{[k]}(n)\textbf{x}^{[k]}(n)\nonumber\\
\!&+&\!\!\!\!\!\!\!\!\!\!\!\sum\limits_{j = 1,j\neq
k}^K\!\!\!\!\!\!\textbf{U}^{[k]\dagger}\!(n)\textbf{H}^{[kj]}\!(n)\textbf{V}^{[ j]}\!(n)\textbf{x}^{[j]}\!(n)\!+\!\textbf{U}^{[k]\dagger}\!(n)\textbf{z}^{[k]}\!(n),\label{2-1}
\end{eqnarray}
where $\textbf{H}^{[kj]}(n)\in\mathbb{C}^{N^{[k]}\times M^{{[j]}}}$ denotes the
channel gain matrix from transmitter $j$ to receiver $k$ in the $n$th time slot. For a symmetric network, each entry of $\textbf{H}^{[kj]}(n)$ can be assumed to be independent and identically distributed (i.i.d.) $\mathcal{CN}(0, a_p)$, where $0<a_p\leq 1$, which is determined according to the signal attenuation due to path loss. For the convenience of analysis, $a_p$ is assumed to be 0.1 in this paper. Block fading channel is adopted in this paper \cite{blockfading}, and for clarity, the time slot number $n$ is henceforth suppressed unless stated otherwise.
$\textbf{x}^{[k]}$ is composed of $d^{[k]}$ data streams of user $k$ with power constraint $P_t^{[k]}=P_t$ for the symmetric network, except when power allocation is considered in Section V-B.
$\textbf{V}^{[k]}\in\mathbb{C}^{M^{[k]}\times d^{{[k]}}}$ and $\textbf{U}^{[k]}\in\mathbb{C}^{N^{[k]}\times d^{{[k]}}}$ are
the unitary precoding and
interference suppression matrices of user $k$, respectively.
$\textbf{z}^{[k]}\in\mathbb{C}^{N^{[k]}\times 1}$ represents receiver noise vector with distribution $\mathcal{CN}(\textbf{0}, \textbf{I}_{N^{[k]}})$ at receiver $k$.

When IA is feasible \cite{Yetis2010}, the interferences among users can be assumed to be perfectly eliminated if the following conditions are satisfied \cite{Gomadam11}.
\begin{equation}
\textbf{U}^{[k]\dagger}\textbf{H}^{[kj]}\textbf{V}^{[j]}=0,\hspace{1mm}\forall
j\neq k,\label{2-2}
\end{equation}
\begin{equation}
\mathrm{rank}\left(\textbf{U}^{[k]\dagger}\textbf{H}^{[kk]}\textbf{V}^{[k]}\right)=d^{[k]}.\label{2-3}
\end{equation}
 Thus the desired signals of user $k$ can be assumed to be received through a $d^{[k]}\times d^{[k]}$ full rank channel matrix $\overline{\textbf{H}}^{[kk]}\triangleq\textbf{U}^{[k]\dagger}\textbf{H}^{[kk]}\textbf{V}^{[k]}$, and \eqref{2-1} can be simplified as
\begin{equation}
\textbf{y}^{[k]}=\overline{\textbf{H}}^{[kk]}\textbf{x}^{[k]}
+\overline{\textbf{z}}^{[k]},\label{2-4}
\end{equation}
where $\overline{\textbf{z}}^{[k]}=\textbf{U}^{[k]\dagger}\textbf{z}^{[k]}$, following $\mathcal{CN}(\textbf{0}, \textbf{I}_{d^{[k]}})$.

In pursuing the matrices of $\textbf{U}^{[k]}$ and $\textbf{V}^{[k]}$, IA only focuses on condition \eqref{2-2} to eliminate the interferences, and does not involve the direct channel $\textbf{H}^{[kk]}$ to maximize the desired signal power within the desired signal subspace \cite{Minn2013}. Thus several IA algorithms have been proposed to further improve the performance of the conventional IA algorithm \cite{Gomadam11, Santamaria10, Guler11, TWC}.

Based on the above description, the transmission rate of user $k$ in the IA network, if only the ID terminal at receiver $k$ is performed, can be expressed as
\begin{equation}
R^{[k]}= \mathrm{log}_2\left|\textbf{I}_{d^{[k]}}+\frac{P_t}{d^{[k]}}\overline{\textbf{H}}^{[kk]}\overline{\textbf{H}}^{[kk]\dagger}\right|.\label{2-5}
\end{equation}

Since this paper mainly concentrates on the information and power transfer in IA networks instead of degrees of freedom (DoFs), it is assumed that there is only one data stream for each user in the rest of this paper\footnote{The conclusion for the situation with more streams can be easily extended, and it is out of the scope of this paper.}. Besides, symmetric networks are considered, and thus all the users have the same parameters, i.e., $M^{[k]}=M$, $N^{[k]}=N$ and $d^{[k]}=1$ for all $k$. Thus the interference alignment is feasible in this paper when condition \eqref{2-2} is met if
\begin{equation}
M+N\geq K+1.\label{2-6}
\end{equation}

\subsection{Wireless Power Transfer in IA Networks}
Each transmitter of the IA wireless network is assumed to have a constant power source, while all the receivers need to replenish energy from WPT or recharge the batteries when WPT is insufficient.
The received signal with one data stream of user $k$ in the IA network before processed by the interference suppression vector $\textbf{u}^{[k]}$ can be expressed as
\begin{equation}
\widehat{\textbf{y}}^{[k]}=\sum\limits_{j = 1}^K\textbf{H}^{[kj]}\textbf{v}^{[ j]}x^{[j]}+\textbf{z}^{[k]}.\label{2-7}
\end{equation}
In \eqref{2-7}, each element of $\widehat{\textbf{y}}^{[k]}$ is the received signal on the corresponding antenna at receiver $k$. $\textbf{v}^{[ j]}$ and $x^{[j]}$ are the precoding vector and transmitted data stream of user $j$, respectively. Denote
\begin{equation}
\mathbb{E}\left(\left|x^{[j]}\right|^2\right)=P_t\mathbb{E}\left(\left|\xi^{[j]}\right|^2\right)=P_t,\label{2-8}
\end{equation}
where $\xi^{[j]}=\displaystyle\frac{x^{[j]}}{\sqrt{P_t}}$ and $\mathbb{E}\left(\left|\xi^{[j]}\right|^2\right)=1$.

If an EH terminal is also equipped at each receiver of the IA network, energy can be harvested through wireless power transfer. Although the energy is captured in RF band at EH terminals, the harvested energy in RF band is proportional to the corresponding energy from baseband signal due to the law of energy conservation.
 Assume that the received signal $\widehat{\textbf{y}}^{[k]}$ on the antennas of receiver $k$ is intended only for WPT, and the harvested energy due to the background noise is negligible and can be ignored. Thus the instantaneous power harvested in a certain time slot at receiver $k$ when the blocking fading is adopted can be calculated as
\begin{eqnarray}
Q^{[k]}&=&\zeta \left(\sum\limits_{j = 1}^K\textbf{H}^{[kj]}\textbf{v}^{[ j]}x^{[j]}\right)^\dagger\cdot\left(\sum\limits_{j = 1}^K\textbf{H}^{[kj]}\textbf{v}^{[ j]}x^{[j]}\right)\nonumber\\
&=&\zeta\left\|\sum\limits_{j = 1}^K\textbf{H}^{[kj]}\textbf{v}^{[ j]}x^{[j]}\right\|_2^2\nonumber\\
&=&\zeta P_t\left\|\sum\limits_{j = 1}^K\textbf{H}^{[kj]}\textbf{v}^{[ j]}\xi^{[j]}\right\|_2^2,\label{2-9}
\end{eqnarray}
where $\zeta\in(0,1)$ is a constant representing the loss in the energy transducer for converting the harvested energy to electrical energy to be stored \cite{rzhang2013}. $\zeta$ is assumed to be 0.5 throughout this paper for the convenience of analysis.

From the above description, we can see that both ID terminal and EH terminal can be equipped at each receiver to transfer information and power, respectively. Power splitter can be used to induce the received power to ID or EH terminals according to the requirements of the system \cite{rzhang2013}, as shown in Fig. 1. $\rho^{[k]}\in[0, 1]$ is the portion of signal power split to the ID terminal at receiver $k$, and correspondingly $1-\rho^{[k]}$ is that to the EH terminal.
\begin{figure}[tp]\label{fig:1}
\centering
\includegraphics[bb=-2 -15 505 383,scale=.54]{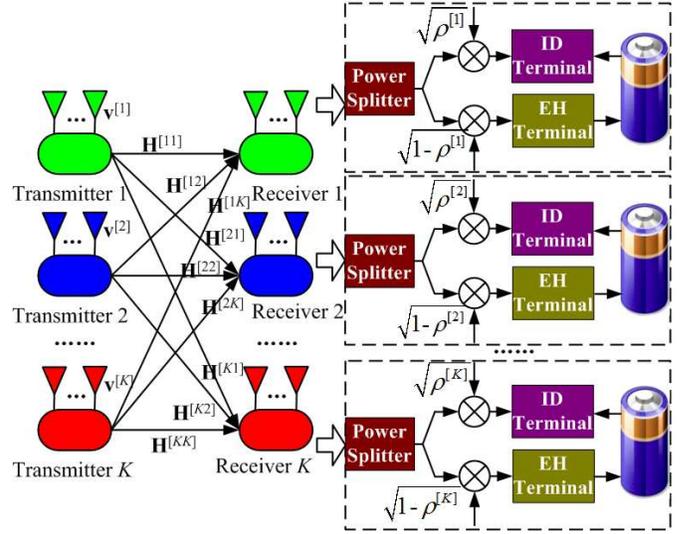}
\caption{A $K$-user IA wireless network with both ID and EH terminals through power splitter at each receiver.}
\end{figure}

\section{Performance Analysis of Simultaneous Wireless Information and Power Transfer in Interference Alignment Networks}
In this section, we first analyze the performance of wireless power transfer in IA networks. Then, we analyze the performance of information transfer in IA networks.
\subsection{Performance Analysis of Wireless Power Transfer in IA Networks}
We analyze performance of energy harvesting in IA networks when a user  is dedicated to wireless power transfer.

\textbf{Lemma 1:} For a given channel gain matrix $\textbf{H}^{[kj]}$ as in \eqref{2-1} and an arbitrary unitary vector $\textbf{v}^{[j]}$, it abides by
\begin{equation}
\left\|\textbf{H}^{[kj]}\textbf{v}^{[j]}\right\|_2\leq \left\|\textbf{H}^{[kj]}\right\|_2=\sqrt{\lambda_{\max}\left(\textbf{H}^{[kj]\dagger}\textbf{H}^{[kj]}\right)}.\nonumber
\end{equation}
\begin{proof}
$\textbf{v}^{[j]}$ is unitary, and thus we have $\left\|\textbf{v}^{[j]}\right\|_2=1$. Based on the definition of the induced norm, it can be obtained that
\begin{eqnarray}\label{effective channel3}
{\left\| {{{{\textbf{H}}}^{[kj]}}} \right\|_2} &=& \mathop {\max }\limits_{{\textbf{v}^{[j]}} \ne {\textbf{0}}} \frac{{{{\left\| {{{{\textbf{H}}}^{[kj]}}{\textbf{v}^{[j]}}} \right\|}_2}}}{{{{\left\| {\textbf{v}^{[j]}} \right\|}_2}}}\nonumber\\
 &=& \mathop {\max }\limits_{{{\left\| {\textbf{v}^{[j]}} \right\|}_2} = 1} {\left\| {{{{\textbf{H}}}^{[kj]}}{\textbf{v}^{[j]}}} \right\|_2}.\label{3-10}
\end{eqnarray}
Thus we have
\begin{equation}
\left\|\textbf{H}^{[kj]}\textbf{v}^{[j]}\right\|_2\leq \left\|\textbf{H}^{[kj]}\right\|_2.\label{3-11}
\end{equation}

According to the definition of spectral norm, we can also obtain
\begin{equation}
\left\|\textbf{H}^{[kj]}\right\|_2=\sqrt{\lambda_{\max}\left(\textbf{H}^{[kj]\dagger}\textbf{H}^{[kj]}\right)}.\label{3-12}
\end{equation}

From \eqref{3-11} and \eqref{3-12}, the conclusion can be obtained as
\begin{equation}
\left\|\textbf{H}^{[kj]}\textbf{v}^{[j]}\right\|_2\leq \left\|\textbf{H}^{[kj]}\right\|_2=\sqrt{\lambda_{\max}\left(\textbf{H}^{[kj]\dagger}\textbf{H}^{[kj]}\right)}.\label{13}
\end{equation}

Thus the upper bound of $\left\|\textbf{H}^{[kj]}\textbf{v}^{[j]}\right\|_2$ can be expressed as $\sqrt{\lambda_{\max}\left(\textbf{H}^{[kj]\dagger}\textbf{H}^{[kj]}\right)}$.
\end{proof}

Based on Lemma 1, Theorem 1 can be derived, which defines the lower and upper bounds of power harvested at receiver $k$ in the IA wireless network.

\textbf{Theorem 1:} In a $K$-user IA wireless network with one data stream for each user, if receiver $k$ is dedicated to WPT utilization, its harvested power should follow
\begin{equation}
0\leq Q^{[k]}\leq \zeta P_t\left(\sum_{j=1}^{K}\sqrt{\lambda_{\max}\left(\textbf{H}^{[kj]\dagger}\textbf{H}^{[kj]}\right)}\right)^2.\nonumber
\end{equation}

\begin{proof}
As demonstrated in \eqref{2-7}, the desired signal received on antennas of receiver $k$ can be expressed as $\mathbf{H}^{[kk]}\textbf{v}^{[k]}x^{[k]}$, and interferences can be denoted as $\mathbf{H}^{[kj]}\textbf{v}^{[j]}x^{[j]}$, $j=1,2,..., K$, $j\neq k$. According to the condition \eqref{2-2} of IA, interferences are constrained into a certain subspace at receiver $k$, which is different from that of the desired signal $\mathbf{H}^{[kk]}\textbf{v}^{[k]}$.

We can assume that the interference from transmitter $j$ to receiver $k$ can be expressed as
\begin{equation}
\textbf{H}^{[kj]}\textbf{v}^{[j]}x^{[j]}=\textbf{g}_j^{[k]},j=1,2,...,K,\hspace{1mm}j\neq k,\label{14}
\end{equation}
which lies in the subspace $\mathcal{Z}^{[k]}$ of $\mathbb{C}^{N\times 1}$ that is orthogonal to $\textbf{u}^{[k]}$ according to \eqref{2-2}, as shown in Fig. 2.
\begin{figure}[tp]
\centering
\includegraphics[bb=-62 -5 507 220,scale=.65]{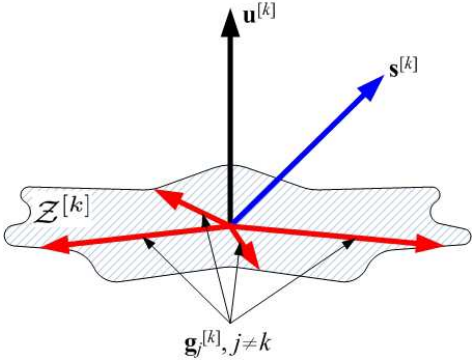}
\caption{Illustration of desired signal, interferences, and $\textbf{u}^{[k]}$ at receiver $k$ in IA networks.}
\end{figure}
Thus the sum of interferences due to all the other transmitters at receiver $k$ can be denoted as
\begin{equation}
\sum_{j=1,j\neq k}^K\textbf{H}^{[kj]}\textbf{v}^{[j]}x^{[j]}=\sum_{j=1,j\neq k}^K \textbf{g}_j^{[k]}=\textbf{g}^{[k]}.\label{15}
\end{equation}
We can also define the desired of user $k$ as
\begin{equation}
\textbf{H}^{[kk]}\textbf{v}^{[k]}x^{[k]}=\textbf{s}^{[k]}.\label{16}
\end{equation}
From Fig. 2, we can see that the interferences at receiver $k$, $\textbf{g}_j^{[k]}$ lying in $\mathcal{Z}^{[k]}$, are orthogonal to $\textbf{u}^{[k]}$, and the desired signal $\textbf{s}^{[k]}$ is randomly distributed in $\mathbb{C}^{N\times 1}$.

From \eqref{2-9}, we can know that
\begin{eqnarray}
Q^{[k]}&=&\zeta\left\|\sum_{j=1}^K\textbf{H}^{[kj]}\textbf{v}^{[j]}x^{[j]}\right\|_2^2\nonumber\\
&=&\zeta\left\|\sum_{j=1,j\neq k}^K \textbf{g}_j^{[k]}+ \textbf{s}^{[k]}  \right\|_2^2\nonumber\\
&=&\zeta\left\|\textbf{g}^{[k]}+ \textbf{s}^{[k]}  \right\|_2^2.\label{17}
\end{eqnarray}
Thus due to the property of vector norm, the following can be achieved
\begin{equation}
0\leq Q^{[k]}\leq\zeta\left(\left\|\textbf{g}^{[k]}\right\|_2+\left\|\textbf{s}^{[k]}\right\|_2  \right)^2.\label{18}
\end{equation}

In \eqref{18}, $Q^{[k]}=0$ when $\textbf{g}^{[k]}=-\textbf{s}^{[k]}$, which means that the sum of interferences and desired signal have the same length with angle $\theta$ (that equals to $\pi$) between them, as illustrated in Fig. 3(a).
\begin{figure}[tp]
\centering
\includegraphics[bb=-80 -2 507 215,scale=.6]{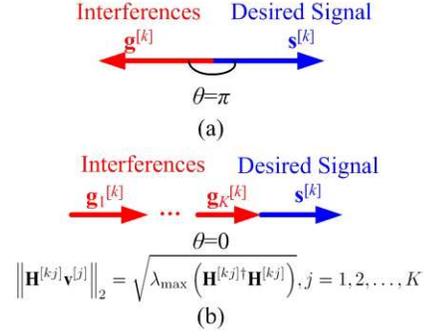}
\caption{Demonstration of lower and upper bounds of power harvested of a certain user in IA networks.}
\end{figure}

From Lemma 1, we can know that
\begin{equation}
\left\|\textbf{g}_j^{[k]}\right\|_2=\left\|\textbf{H}^{[kj]}\textbf{v}^{[j]}x^{[j]}\right\|_2\leq \sqrt{P_t\cdot\lambda_{\max}\left(\textbf{H}^{[kj]\dagger}\textbf{H}^{[kj]}\right)},\label{19}
\end{equation}
where $j=1,2,...,K,\hspace{1mm}j\neq k$, and
\begin{equation}
\left\|\textbf{s}^{[k]}\right\|_2=\left\|\textbf{H}^{[kk]}\textbf{v}^{[k]}x^{[k]}\right\|_2\leq \sqrt{P_t\cdot\lambda_{\max}\left(\textbf{H}^{[kk]\dagger}\textbf{H}^{[kk]}\right)}.\label{20}
\end{equation}

From \eqref{19} and \eqref{20}, it can also be obtained that
\begin{eqnarray}
&&\zeta\left(\left\|\textbf{g}^{[k]}\right\|_2+\left\|\textbf{s}^{[k]}\right\|_2  \right)^2\nonumber\\
&=&\zeta\left(\left\|\sum_{j=1,j\neq k}^K \textbf{g}_j^{[k]}\right\|_2+\left\|\textbf{s}^{[k]}\right\|_2\right)^2\nonumber\\
&\leq&\zeta\left(\sum_{j=1,j\neq k}^K\left\| \textbf{g}_j^{[k]}\right\|_2+\left\|\textbf{s}^{[k]}\right\|_2\right)^2\nonumber\\
&\leq&\zeta P_t\left(\sum_{j=1}^{K}\sqrt{\lambda_{\max}\left(\textbf{H}^{[kj]\dagger}\textbf{H}^{[kj]}\right)}\right)^2.\label{21}
\end{eqnarray}

The upper bound of $Q^{[k]}$ in \eqref{21} can be achieved only when all the interferences and desired signal are all in the same direction, i.e., $\theta=0$, and every vector $\textbf{v}^{[j]}$ can make $\left\|\textbf{H}^{[kj]}\textbf{v}^{[j]}\right\|_2$ achieve the largest value as in \eqref{13}, which is illustrated in Fig. 3(b).

From \eqref{18} and \eqref{21}, we can have the following conclusion
\begin{equation}
0\leq Q^{[k]}\leq \zeta P_t\left(\sum_{j=1}^{K}\sqrt{\lambda_{\max}\left(\textbf{H}^{[kj]\dagger}\textbf{H}^{[kj]}\right)}\right)^2.\label{22}
\end{equation}
The lower and upper bounds of power harvested at the $k$th receiver $Q^{[k]}$ can be denoted as 0 and $\zeta P_t\left(\sum_{j=1}^{K}\sqrt{\lambda_{\max}\left(\textbf{H}^{[kj]\dagger}\textbf{H}^{[kj]}\right)}\right)^2$, respectively.
\end{proof}

\emph{Remark 1:} The lower bound of $Q^{[k]}$ in \eqref{22}, 0,  can be approached in some certain cases, while its upper bound is difficult to achieve due to the following reasons.
\begin{itemize}
\item It is difficult for all the interferences $\textbf{g}_j^{[k]}$, $j\neq k$, and desired signal $\textbf{s}^{[k]}$ to lie in the same direction, which means the angle between these vectors should be 0.
\item Each vector $\textbf{v}^{[j]}$ of transmitter $j$ is designed according to the requirement of IA in \eqref{2-2}, instead of intending to achieve the goal in \eqref{13}. Thus it is difficult to require all the precoding vectors to achieve the upper bound of $\left\|\textbf{H}^{[kj]}\textbf{v}^{[j]}\right\|_2$ in \eqref{13}.
\end{itemize}

Fig. 4 shows the upper bound of $Q^{[k]}$, which is defined in \eqref{22},  compared with its simulated value in a 5-user IA network over 10000 time slots. From Fig. 4, we can see that, with a low probability, the simulated value of $Q^{[k]}$ can approach its upper bound described in \eqref{22}. On the other hand, the simulated value of $Q^{[k]}$ can get close to its lower bound, 0, with a much larger probability. Thus the results in Fig. 4 are consistent with \emph{Remark 1}.
\begin{figure}[t]
\centering
\includegraphics[bb=150 300 507 540,scale=.85]{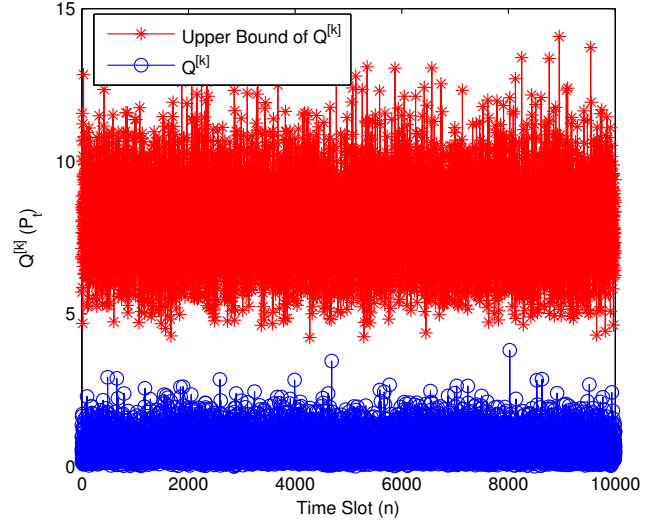}
\caption{Comparison of upper bound and simulated value of $Q^{[k]}$ in a 5-user IA network with 1 data stream each user.}
\end{figure}

\subsection{Performance of Information Transmission of IA}
When a user is dedicated to information transmission in the IA network, its performance will also be varying according to the CSI of the network. In our previous work \cite{TWC}, it is proved that the information transmission performance of user $k$ in the IA network with one data stream each user is determined by the length of the desired signal $c_k$, and  the angle $\delta_k$ between the directions of interference suppression vector $\textbf{u}^{[k]}$ and desired signal. The transmission rate of user $k$ can be expressed as
\begin{eqnarray}
R^{[k]}&=&\mathrm{log}_2\left(1+P_t\left|\textbf{u}^{[k]\dagger}\textbf{H}^{[kk]}\textbf{v}^{[k]}\right|^2\right)\nonumber\\
&=&\mathrm{log}_2\left(1+P_tc_k^2\cos^2\delta_k\right),\label{IArate}
\end{eqnarray}
where $c_k=\left|\textbf{H}^{[kk]}\textbf{v}^{[k]}\right|$ and $\delta_k$ is the angle between $\textbf{u}^{[k]}$ and $\textbf{H}^{[kk]}\textbf{v}^{[k]}$.

In Fig. 5, the transmission rate of user $k$ when receiver $k$ is used as an ID terminal, $R^{[k]}$, and the power harvested of user $k$ when receiver $k$ is EH terminal, $Q^{[k]}$, are compared in a 5-user IA network over 50 time slots, when the average received SNR is 10dB. The average received SNR for user $k$ can be expressed as $10\lg \left(P_t\mathbb{E}\left(\left|\textbf{u}^{[k]\dagger}\textbf{H}^{[kk]}\textbf{v}^{[k]}\right|^2\right)\right)$.
\begin{figure}[t]
\centering
\includegraphics[bb=147 293 490 539,scale=.85]{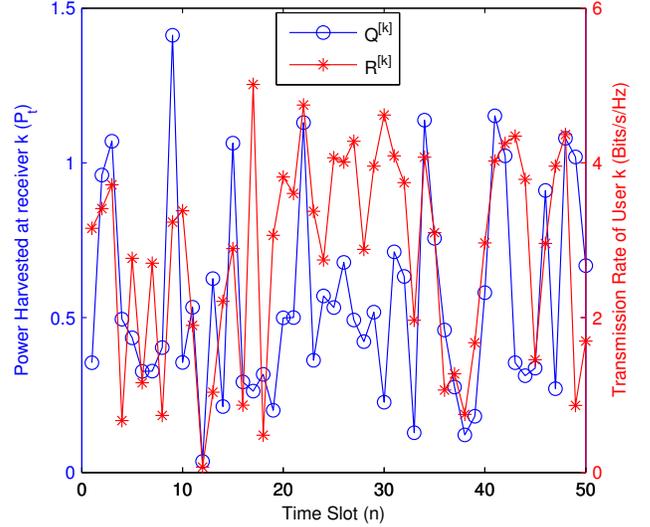}
\caption{Comparison of $R^{[k]}$ and $Q^{[k]}$ in a 5-user IA network with 1 data stream each user, when the average received SNR is 10dB.}
\end{figure}

From \cite{TWC} and Theorem 1, we can see that, for the CSI in a time slot, the performance of information transmission and that of WPT may be quite different, as shown in Fig. 5. Thus we should determine carefully to which extent a receiver is selected to harvest energy according to the difference between the performance of information transmission and that of WPT.

\section{SWIPT-User Selection Scheme in IA Networks}
In Fig. 1, one power splitter is equipped at each receiver; however, it is improper to still adopt power splitters due to the limitation of the size and complexity of the receivers in some practical systems.
In this section, a SWIPT-user selection scheme is proposed for simultaneous wireless information and power transfer in IA networks. In the SWIPT-US scheme, a number of receivers are selected for EH dedicatedly, which are similar to those with $\rho=0$ in Fig. 1, while the other receivers are devoted to ID, which are similar to those with $\rho=1$. Thus the SWIPT-US scheme is much easier to be implemented than the one with power splitters. Two algorithms to realize the SWIPT-US scheme are proposed.

\subsection{Round-Robin Selection Algorithm for SWIPT-US}
In the IA wireless network, every receiver can act as an EH or ID terminal in a time slot, and simultaneous wireless information and power transfer can be achieved. In practical systems, we should not adopt all the receivers as EH terminals, because information transmission of the network should not be terminated; on the other hand, the receivers should not be all dedicated to ID either, as the receivers need to collect energy to support their operation, and prolong the life time of their batteries. Thus, only a fraction of the receivers should be devoted to EH in a time slot, and the SWIPT-US scheme for SWIPT-based IA networks can be leveraged.

A simple idea of SWIPT-US is to select a number of receivers as EH terminals in a round-robin principle, i.e., to assign SWIPT users in successive time slots in circular order without priority, which is called round-robin selection (RRS) algorithm for the SWIPT-US scheme. Assume that $L$ receivers are switched to EH terminals in the SWIPT-US scheme, $L<K$. In a $K$-user IA network with both ID and EH terminals at each receiver as shown in Fig. 1, RRS algorithm for SWIPT-US can be represented by the following steps:
\begin{enumerate}
\item When time slot $n$ starts, the solutions of IA are calculated through using the minimizing interference leakage (MinIL) IA algorithms \cite{Gomadam11}.
\item The largest user number of dedicated EH users in time slot $n-1$ is $\mathcal{I}_{n-1}$, and thus the users with number $(\mathcal{I}_{n-1}+1)\%K,(\mathcal{I}_{n-1}+2)\%K,\dots,(\mathcal{I}_{n-1}+L)\%K$ are selected as EH users in time slot $n$ (user 0 is equal to user $K$). The other $K-L$ users are assigned to be ID users.
\item Set $\mathcal{I}_{n}=(\mathcal{I}_{n-1}+L)\%K$.
\item The portion of signal power split to the ID terminal, $\rho$, is set to be 0 at the selected $L$ EH receivers, and $\rho$ is set to 1 at all the other ID receivers.
\item Information transmission begins at the $K-L$ ID receivers according to \eqref{IArate}, and energy harvesting is also performed at the $L$ EH receivers according to \eqref{2-9}.
\item  After duration $\mathcal{T}$, time slot $n$ ends, $n=n+1$, and go back to Step 1).
\end{enumerate}

The RRS algorithm for SWIPT-US is simple to implement. Nevertheless, both of the energy harvested and the transmission rate of the IA network can be further improved with the same number of EH users $L$.

\subsection{PRR-Based Selection Algorithm for SWIPT-US}
In order to further improve  EH and ID efficiency of the IA network, we define a parameter called power-to-rate ratio (PRR) to compare the instantaneous EH capability to ID capability of an IA user. The PRR of user $k$ in the IA network in a time slot can be denoted as
\begin{equation}
\eta^{[k]}\!=\!\frac{Q^{[k]}}{R^{[k]}}\!=\!\frac{\zeta P_t\left\|\sum\limits_{j = 1}^K\textbf{H}^{[kj]}\textbf{v}^{[ j]}\xi^{[j]}\right\|_2^2}{\mathrm{log}_2\!\left(1+P_t\left|\textbf{u}^{[k]\dagger}\textbf{H}^{[kk]}\textbf{v}^{[k]}\right|^2\right)
}\label{23}
\end{equation}

The PRR of user $k$, $\eta^{[k]}$, defines the ratio between the power transferred if user $k$ is only assigned as an EH terminal and the information rate if it is specially selected as an ID terminal. When the PRR of user $k$ is large, it means that it is better for user $k$ to harvest energy than to transmit information. With PRR, we propose a PRR-based selection (PRRS) algorithm for SWIPT-US, in which the users with larger PRR are selected as EH receivers in a time slot. The PRRS algorithm for SWIPT-US can be expressed by the following steps:
\begin{enumerate}
\item When time slot $n$ starts, the solutions of IA are calculated through using MinIL IA algorithms.
\item The PRRs of all the users are calculated according to \eqref{23}, and we can obtain $\eta^{[1]},\eta^{[2]},\dots,\eta^{[K]}$.
\item Select $L$ users with the largest PRRs as EH receivers, and devote the other users to ID.
\item The portion of signal power split to the ID terminal, $\rho$, is set to be 0 at the selected $L$ EH receivers, and $\rho$ is set to 1 at all the other ID receivers.
\item Information transmission begins at the $K-L$ ID receivers according to \eqref{IArate}, and energy harvesting is also performed at the $L$ EH receivers according to \eqref{2-9}.
\item  After duration $\mathcal{T}$, time slot $n$ ends, $n=n+1$, and go back to Step 1).
\end{enumerate}

\emph{Remark 2:} Comparing RRS and PRRS algorithms for SWIPT-US, we can have the following observations.
\begin{itemize}
\item The RRS algorithm is easy to implement, as it adopts round-robin principle to select EH receivers, and no additional calculation is needed. By contrast, the PRRS algorithm is more complex, because it needs to calculate the PRR parameter of all the users to select EH receivers.
\item Both EH and ID performance of the PRRS algorithm is better than that of the RRS algorithm with the same number of EH receivers, due to the selection according to the PRR parameter in the PRRS algorithm.
\item In both RRS and PRRS algorithms, a receiver may be dedicated to be either an EH terminal or an ID terminal, and the EH and ID performance of the network may not be continuously optimized according to the requirements of the systems, e.g., battery status. In addition,  they do not consider the specific requirements of rate and energy for the users. Thus, if power splitters can be equipped at the receivers, the EH and ID performance of the network can be improved significantly, and the requirements of the users can be better satisfied. Nevertheless, the SWIPT-US scheme is much easier to be implemented.
\end{itemize}

\section{Optimization of both ID and EH Performance Using Power Splitting}
In Section IV, SWIPT-US algorithms for IA networks are proposed, and they are simple and easy-implemented. However, the EH and ID performance may not be continuously optimized, and the specific requirements of the users are not considered. Thus in this section, a power-splitting optimization algorithm is proposed, when power splitters are available, to optimize the performance of the network over the portion of signal power splitting $\rho$. Transmitted power allocation among users in the PSO algorithm is also studied.
\subsection{Power-Splitting Optimization for SWIPT in IA Networks}
 In this section, it is assumed that each receiver in the IA network can serve as EH and ID terminals simultaneously according to Fig. 1. $\rho^{[k]}$ is no long 1 or 0 for user $k$, instead, the received power is split into two portions, which are induced to EH and ID terminals, respectively. According to $\rho^{[k]}, k=1,2,\dots,K$, the sum rate of the IA network can be represented as
 \begin{eqnarray}
 \!\!\!\!\!\!\!\!\mathcal{SR}\!\!\!\!\!&=&\!\!\!\!\!\sum_{k=1}^{K}\mathcal{R}_{\rho^{[k]}}^{[k]}\nonumber\\
 &=&\!\!\!\!\!\sum_{k=1}^{K}\mathrm{log}_2\!\left(1\!+\!\rho^{[k]}P_t\left|\textbf{u}^{[k]\dagger}\textbf{H}^{[kk]}\textbf{v}^{[k]}\right|^2\right)\!.\label{24}
 \end{eqnarray}
 The sum power harvested in the IA network should also be updated as
 \begin{eqnarray}
\mathcal{SQ}&=&\sum_{k=1}^{K}\mathcal{Q}_{\rho^{[k]}}^{[k]}\nonumber\\
&=&\sum_{k=1}^{K}\left(1-\rho^{[k]}\right)\zeta P_t\left\|\sum\limits_{j = 1}^K\textbf{H}^{[kj]}\textbf{v}^{[ j]}\xi^{[j]}\right\|_2^2.\label{25}
\end{eqnarray}

Thus the optimization problem of the PSO algorithm considering both EH and ID performance simultaneously at each receiver can be defined as \eqref{26} (on the next page).

\begin{figure*}
\begin{eqnarray}
\max_{\rho^{[1]},\rho^{[2]},\dots,\rho^{[K]}} \!\!\!\!\!&\displaystyle\sum\limits_{k = 1}^K&\!\!\!\! \!\left(\alpha^{[k]}\mathrm{log}_2\!\left(1+\rho^{[k]}P_t\left|\textbf{u}^{[k]\dagger}\textbf{H}^{[kk]}\textbf{v}^{[k]}\right|^2\right) +\beta^{[k]}\left(1-\rho^{[k]}\right)\zeta P_t\left\|\sum\limits_{j = 1}^K\textbf{H}^{[kj]}\textbf{v}^{[ j]}\xi^{[j]}\right\|_2^2\right)=\mathcal{F}\nonumber\\
s.t.\hspace{8mm} \!\!\!\!\!\!&0&\!\!\!\!\!\! \leq \rho^{[k]}\leq 1,\hspace{1mm}\forall k=1,2,\dots,K.\label{26}
\end{eqnarray}
\end{figure*}

In \eqref{26}, $\alpha^{[k]}$ and $\beta^{[k]}$ are two nonnegative design parameters, which denote the weights for the requirements of rate and energy needed of user $k$, respectively, and $\alpha^{[k]}+\beta^{[k]}=1$. When $\alpha^{[k]}$ becomes larger, it means that the transmission-rate requirement of user $k$ is high or the battery power of receiver $k$ is sufficient; when $\alpha^{[k]}$ is smaller, it means that the rate requirement of user $k$ is low or the battery is running out.

For example, we can define $\alpha^{[k]}$ and $\beta^{[k]}$ as
 \begin{equation}
\alpha^{[k]}=\frac{\upsilon^{[k]}\cdot R_{req}^{[k]}}{\upsilon^{[k]}\cdot R_{req}^{[k]}+\varphi^{[k]}\cdot Q_{req}^{[k]}},\label{27}
\end{equation}
 \begin{equation}
\beta^{[k]}=\frac{\varphi^{[k]}\cdot Q_{req}^{[k]}}{\upsilon^{[k]}\cdot R_{req}^{[k]}+\varphi^{[k]}\cdot Q_{req}^{[k]}},\label{28}
\end{equation}
where $R_{req}^{[k]}$ and $Q_{req}^{[k]}$ are instantaneous requested transmission rate and requested power by user $k$. $\upsilon^{[k]}$ and $\varphi^{[k]}$ are two constants to make rate correspond to power to achieve proper values of $\alpha^{[k]}$ and $\beta^{[k]}$.

The optimization problem in \eqref{26} is a convex optimization problem, and its optimal solutions can be calculated as in Theorem 2.

\textbf{Theorem 2:} The optimization problem in \eqref{26} is a convex optimization problem in each time slot, and $\forall k\in \{1,2,\dots,K\}$, its closed-form optimal solutions can be expressed as
\begin{equation}
\rho^{*[k]}=\left\{\begin{array}{ll}
                  0,\hspace{10mm}\psi^{[k]}\leq 0\nonumber\\
                  1,\hspace{10mm}\psi^{[k]}\geq 1\nonumber\\
                  \psi^{[k]},\hspace{6.5mm}\textrm{otherwise},\end{array} \right.\nonumber
\end{equation}
where
\begin{equation}
\psi^{[k]}=\displaystyle{\frac{\alpha^{[k]}}{\beta^{[k]}\zeta P_t\left\|\sum\limits_{j = 1}^K\textbf{H}^{[kj]}\textbf{v}^{[ j]}\xi^{[j]}\right\|_2^2\ln2}-\frac{1}{P_t\left|\textbf{u}^{[k]\dagger}\textbf{H}^{[kk]}\textbf{v}^{[k]}\right|^2}}\nonumber.
\end{equation}

\begin{proof}
In a time slot, the solutions of the IA network are determined by the channel state information of the network, and thus $\rho^{[k]}, k=1,2,\dots,K$, are the only variables in \eqref{26}. $\forall k\in\{1,2,\dots,K\}$, $\rho^{[k]}\in[0,1]$ in \eqref{26} is a convex set. It can also easily obtained that
\begin{equation}
\displaystyle{\frac{\partial^2 \mathcal{F}}{\partial \rho^{[k]2}}<0},\label{29}
\end{equation}
and the objective function $\mathcal{F}$ of \eqref{26} is convex. Thus the optimization problem of \eqref{26} is a convex optimization problem.

\begin{figure*}
\begin{eqnarray}
\frac{\partial \mathcal{F}}{\partial \rho^{[k]}}=\frac{\alpha^{[k]}P_t\left|\textbf{u}^{[k]\dagger}\textbf{H}^{[kk]}\textbf{v}^{[k]}\right|^2}{\left(1+P_t\left|\textbf{u}^{[k]\dagger}\textbf{H}^{[kk]}\textbf{v}^{[k]}\right|^2\rho^{[k]}\right)\ln2}-\beta^{[k]}\zeta P_t\left\|\sum\limits_{j = 1}^K\textbf{H}^{[kj]}\textbf{v}^{[ j]}\xi^{[j]}\right\|_2^2=0.\label{30}
\end{eqnarray}
\end{figure*}

We can obtain the derivative of $\mathcal{F}$ with $\rho^{[k]}$ as \eqref{30} (on the next page). Because the solutions to \eqref{26} should satisfy that $0\leq\rho^{[k]}\leq 1$, thus from \eqref{30}, we can achieve the closed-form optimal solutions of \eqref{26} as
\begin{equation}
\rho^{*[k]}=\left\{\begin{array}{ll}
                  0,\hspace{10mm}\psi^{[k]}\leq 0\\
                  1,\hspace{10mm}\psi^{[k]}\geq 1\\
                  \psi^{[k]},\hspace{6.5mm}\textrm{otherwise},\end{array} \right.\label{31}
\end{equation}
where
\begin{equation}
\psi^{[k]}=\displaystyle{\frac{\alpha^{[k]}}{\beta^{[k]}\zeta P_t\left\|\sum\limits_{j = 1}^K\textbf{H}^{[kj]}\textbf{v}^{[ j]}\xi^{[j]}\right\|_2^2\ln2}-\frac{1}{P_t\left|\textbf{u}^{[k]\dagger}\textbf{H}^{[kk]}\textbf{v}^{[k]}\right|^2}}.\label{32}
\end{equation}
\end{proof}

Equations \eqref{31} and \eqref{32} present the easy-implemented closed-from solutions to \eqref{26}, and thus the EH and ID performance of the IA network can be optimized simultaneously according to the specific requirements of the users represented by $\alpha^{[k]}$ and $\beta^{[k]}$, $k=1,2,\dots,K$.

\emph{Remark 3:} From \eqref{31} and \eqref{32}, we can find that the closed-form optimal solution $\rho^{*[k]}$ of user $k$ is not affected by $\alpha$ and $\beta$ parameters of other users, i.e., the specific power and rate requirements of the users in the network will not interact among users.

\subsection{PSO Algorithm with Power Allocation}
In the above discussions, we assume that equal transmitted power is allocated to each user, $P_t^{[k]}=P_t, k=1,2,\dots,K$, i.e., power allocation is not involved. In practical systems, the channel is usually not symmetric, and power allocation should be considered to guarantee the performance of the whole network. Thus in this subsection, power allocation is studied in  PSO to further improve the ID and EH performance of SWIPT in IA networks.

We assume that the sum transmitted power of all the users is constrained to be lower than a constant, i.e., $\sum_{k=1}^{K}P_t^{[k]}\leq K\cdot P_t$, and thus $P_t$ is the average of the transmitted power of each user. When power allocation is considered, the optimization problem in \eqref{26} for the PSO algorithm can be updated as \eqref{33} (on the next page).

\begin{figure*}
\begin{eqnarray}
\max_{\rho^{[1]},\rho^{[2]},\dots,\rho^{[K]},P_t^{[1]},P_t^{[2]},\dots,P_t^{[K]}} \!\!\!\!\!&\displaystyle\sum\limits_{k = 1}^K&\!\!\!\! \!\left(\alpha^{[k]}\mathrm{log}_2\!\left(1+\rho^{[k]}P_t^{[k]}\left|\textbf{u}^{[k]\dagger}\textbf{H}^{[kk]}\textbf{v}^{[k]}\right|^2\right) +\beta^{[k]}\left(1-\rho^{[k]}\right)\zeta\left\|\sum\limits_{j = 1}^K \sqrt{P_t^{[j]}}\textbf{H}^{[kj]}\textbf{v}^{[ j]}\xi^{[j]}\right\|_2^2\right)\nonumber\\
s.t.\hspace{8mm} \!\!\!\!\!\!&0&\!\!\!\!\!\!\! \leq \rho^{[k]}\leq 1,\hspace{1mm}\forall k=1,2,\dots,K,\nonumber\\
&\sum\limits_{k=1}^K&\!\!\!\!P_t^{[k]}\leq K\cdot P_t,\nonumber\\
&P&\!\!\!\!\!\!\!\!_t^{[k]}\geq 0,\hspace{1mm}\forall k=1,2,\dots,K.\label{33}
\end{eqnarray}
\end{figure*}

The optimization problem in \eqref{33} is not convex due to the product of $\rho^{[k]}$ and $P_t^{[k]}$, and thus the closed-form optimal solutions are difficult to obtain. There are many simple but effective methods for solving continuous optimization problems. In the simulations of this paper, the interior-point method \cite{convexbook} is adopted.

\emph{Remark 4:} The power allocation in the PSO algorithm with two extremes, $\forall k$, $\alpha^{[k]}=1$ and  $\alpha^{[k]}=0$, is interesting and worth noting. When $\alpha^{[k]}=1, k=1,2,\dots,K$, only ID terminals are active at the receivers, and \eqref{33} becomes a conventional power allocation problem that can be solved by the standard ``water-filling" power allocation strategy.

When $\alpha^{[k]}=1, k=1,2,\dots,K$, \eqref{33} becomes
\begin{eqnarray}
\max_{P_t^{[1]},P_t^{[2]},\dots,P_t^{[K]}} \!\!\!\!\!&\sum\limits_{k = 1}^K&\!\!\!\! \!\mathrm{log}_2\!\left(1+P_t^{[k]}\left|\textbf{u}^{[k]\dagger}\textbf{H}^{[kk]}\textbf{v}^{[k]}\right|^2\right) \nonumber\\
s.t.\hspace{8mm}&\sum\limits_{k=1}^K&\!\!\!\!P_t^{[k]}\leq K\cdot P_t,\nonumber\\
 &P&\!\!\!\!\!\!\!\!_t^{[k]}\geq 0.\label{34}
\end{eqnarray}
The closed-form solutions to \eqref{34} can be easily achieved by ``water-filling" as \cite{Tse2005book}
\begin{equation}
P_{t\_opt}^{[k]}=\left(\mathcal{V}-\frac{1}{\left|\textbf{u}^{[k]\dagger}\textbf{H}^{[kk]}\textbf{v}^{[k]}\right|^2}\right)^+,\label{35}
\end{equation}
where $x^+\triangleq \max(x,0)$, and $\mathcal{V}$ should satisfy
\begin{equation}
\sum\limits_{k = 1}^{K}\left(\mathcal{V}-\frac{1}{\left|\textbf{u}^{[k]\dagger}\textbf{H}^{[kk]}\textbf{v}^{[k]}\right|^2}\right)^+=K\cdot P_t.\label{36}
\end{equation}

\emph{Remark 5:} When $\alpha^{[k]}=0, k=1,2,\dots,K$, $\rho^{[k]}$ is equal to 0, $\forall k\in \{1,2,\dots,K\}$, and all the receivers work as EH terminals. This can happen when the batteries at receivers are all at low levels and need to be recharged, and \eqref{33} becomes
\begin{eqnarray}
\max_{P_t^{[1]},P_t^{[2]},\dots,P_t^{[K]}} \!\!\!\!\!&\displaystyle\sum\limits_{k = 1}^K&\!\!\!\! \!\left\|\sum\limits_{j = 1}^K\sqrt{P_t^{[j]}}\textbf{H}^{[kj]}\textbf{v}^{[ j]}\xi^{[j]}\right\|_2^2\nonumber\\
s.t.\hspace{8mm}
&\sum\limits_{k=1}^K&\!\!\!\!P_t^{[k]}\leq K\cdot P_t,\nonumber\\
 \!\!\!\!\!\!\!\!\!\!\!&P&\!\!\!\!\!\!\!\!_t^{[k]}\geq 0.\label{37}
\end{eqnarray}

In practical systems, the extreme cases when all the $\alpha^{[k]}$ equal to 1 or 0 will appear with low probability. The common situation is that some receivers with low level battery will harvest more energy with low transmission rate, while some others may have sufficient power supply, and they want to transmit more information instead of energy harvesting.

\section{Simulation Results and Discussions}
We consider a $5$-user IA wireless network with one data stream for each user. Three antennas are equipped at each transceiver (i.e., $M=N=3$).
Rayleigh block channel fading \cite{blockfading} is adopted, and perfect CSI is available at each node. $a_p$ due to the path loss is  0.1 (i.e., each element in the channel matrix follows $\mathcal{CN}(0, 0.1)$), and  $\zeta$ is 0.5 throughout the simulations.

Fig. 6 shows the sum harvested power and sum rate of the 5-user IA network  using the PRRS and RRS algorithms for the SWIPT-US scheme with different numbers of dedicated ID receivers, when the average received SNR is 10dB.
\begin{figure}[t]
\centering
\includegraphics[bb=150 294 507 541,scale=.85]{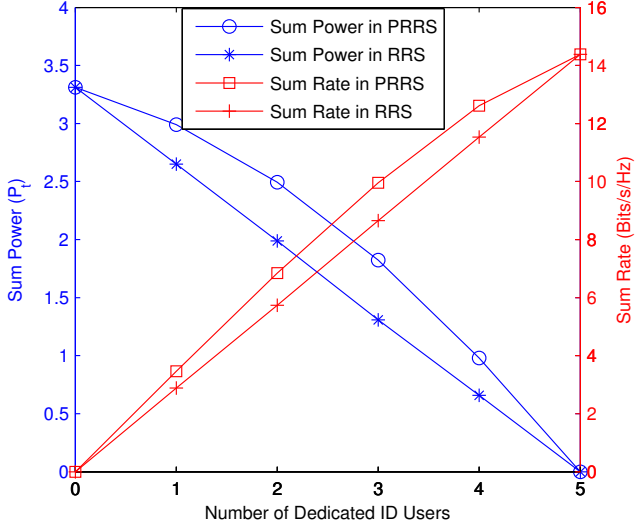}
\caption{Comparison of sum harvested power and sum rate of the 5-user IA network  using the PRRS and RRS algorithms with different numbers of dedicated ID receivers, when the average received SNR is 10dB.}
\end{figure}
From the results, we can see that the sum harvested power increases and the sum rate decreases as the number of dedicated ID users becomes smaller. Both ID performance and EH performance of the PRRS algorithm are better than or at least equal to those of the RRS algorithm due to the selection according to PRR in \eqref{23}. For example, when there are 3 dedicated ID users, the sum power harvested increases from 1.34$P_t$ with the RRS algorithm to 1.83$P_t$ with the PRRS algorithm, and the sum rate increases from 8.63 bits/s/Hz with the RRS algorithm to 9.94 bits/s/Hz with the PRRS algorithm.

In the PSO algorithm, $\alpha^{[k]}$ is a key parameter, which determines the trade-off between the rate and energy performance of user $k$. Assume $\alpha^{[1]}=\alpha^{[2]}=\dots=\alpha^{[K]}=\alpha$, Fig. 7 shows the sum harvested power  and sum rate in a 5-user IA network with different values of $\alpha$, when the average received SNR is 10dB. From the results, we can observe that, when $\alpha$ becomes larger in the PSO algorithm, the sum rate of the network increases, and the sum harvested power of the network decreases. From \eqref{31}, \eqref{32} and $\alpha^{[k]}+\beta^{[k]}=1$, we can know that, when
 \begin{equation}
 \alpha^{[k]}\leq\frac{\zeta \left\|\sum\limits_{j = 1}^K\textbf{H}^{[kj]}\textbf{v}^{[ j]}\xi^{[j]}\right\|_2^2\ln2}{\zeta \left\|\sum\limits_{j = 1}^K\textbf{H}^{[kj]}\textbf{v}^{[ j]}\xi^{[j]}\right\|_2^2\ln2+\left|\textbf{u}^{[k]\dagger}\textbf{H}^{[kk]}\textbf{v}^{[k]}\right|^2},\label{38}
\end{equation}
the optimal solution $\rho^{*[k]}$ will be 0, and receiver $k$ will be dedicated to EH. From Fig. 7, we can also find that, when $ \alpha^{[k]}= \alpha, \forall k\in \{1,2,\dots,K\}$, is set below 0.4, the sum rate of the network will be 0, and all the receivers are adopted as EH terminals. Thus, $ \alpha^{[k]}= \alpha, \forall k\in \{1,2,\dots,K\}$, can be set in a reduced domain of $[0.4 ,1]$.

\begin{figure}[t]
\centering
\includegraphics[bb=150 295 507 543,scale=.85]{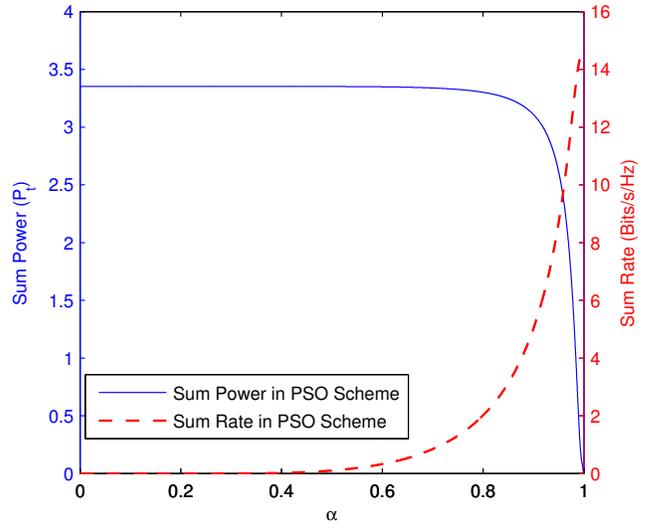}
\caption{Comparison of sum harvested power and sum rate of the PSO algorithm with different values of $\alpha$ in 5-user IA network, when the average received SNR is  10dB.}
\end{figure}

The PSO algorithm can optimize EH and ID performance of IA networks simultaneously, and the performance with different power and rate requirements of users should be discussed. In Fig. 8, we compare the average performance of power harvested, rate transmitted, and corresponding parameter $\rho$ of the users in a 5-user IA network with different values of $\alpha$, when the average received SNR is 10dB. $\alpha^{[1]}=0.6$, $\alpha^{[2]}=0.8$, $\alpha^{[3]}=0.95$, $\alpha^{[4]}=0.975$, $\alpha^{[5]}=0.99$. From the results, it is shown that the different power and rate requirements of the users can be traded off with different values of $\alpha$ in the  PSO algorithm for SWIPT in IA networks. Besides, when the value of $\alpha$ of a user becomes larger, its transmission rate becomes larger, its harvested power becomes smaller, and its corresponding $\rho$ becomes larger.

\begin{figure}[t]
\centering
\includegraphics[bb=150 298 507 550,scale=.85]{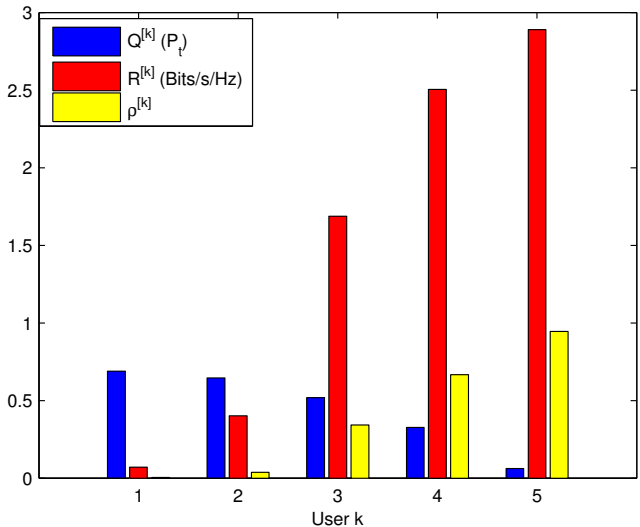}
\caption{The average performance comparison of the power harvested, rate transmitted, and corresponding parameter $\rho$ of users in the PSO algorithm with different values of $\alpha$ of users, when the average received SNR is 10dB. $\alpha^{[1]}=0.6$, $\alpha^{[2]}=0.8$, $\alpha^{[3]}=0.95$, $\alpha^{[4]}=0.975$, $\alpha^{[5]}=0.99$.}
\end{figure}

Power allocation can improve both EH and ID performance in the PSO algorithm of IA networks significantly. Power-rate region can characterize all the achievable power and rate pairs under a given transmit power constraint. Fig. 9 shows the power-rate trade-offs of the PSO algorithm with PA, PSO algorithm without PA, PRRS and RRS algorithms for the SWIPT-US scheme in the 5-user IA network with the same values of $\alpha$ of all the users, i.e, $\alpha^{[1]}=\alpha^{[2]}=\dots=\alpha^{[5]}$. The average received SNR is  10dB, and the average received SNR with power allocation can be calculated as $10\lg \mathbb{E}\left(\sum_{k=1}^{K}\left(P_t^{[k]}\left|\textbf{u}^{[k]\dagger}\textbf{H}^{[kk]}\textbf{v}^{[k]}\right|^2\right)/K\right)$.
\begin{figure}[t]
\centering
\includegraphics[bb=151 288 507 550,scale=.85]{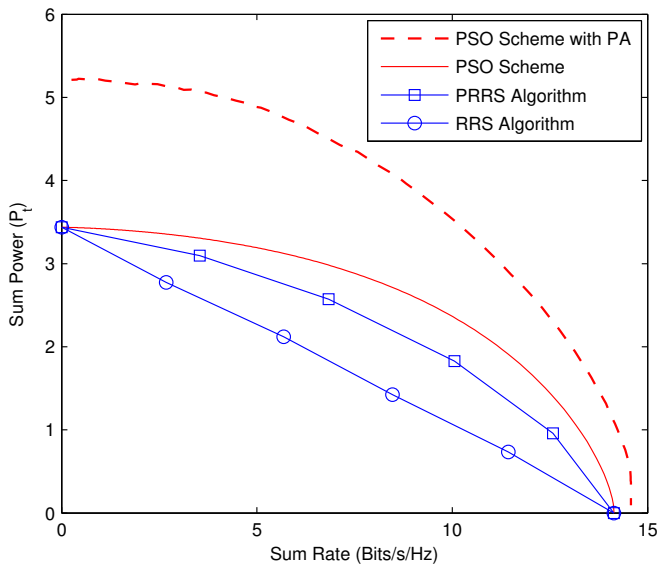}
\caption{Power-rate tradeoffs of the PSO algorithm with PA, PSO algorithm without PA, and PRRS and RRS algorithms for the SWIPT-US scheme, when the average received SNR is 10dB.}
\end{figure}
From the results in Fig. 9, we can observe that power-rate performance of the PSO algorithm is better than that of the SWIPT-US scheme. When PA is adopted in the PSO algorithm, its power-rate performance can be significantly improved, especially the performance of EH, which is consistent with \emph{Remark 4} and \emph{Remark 5}. The enhancement of the EH performance is much more obvious than that of the ID performance by PA in the PSO algorithm, because log function is used in calculation of the rate, while sum of the desired signal and all the interferences is adopted to calculate the power harvested.

In Fig. 10, we show the average power harvested, transmission rate, transmitted power allocated, and corresponding  parameter $\rho$ of the users in the power-allocation PSO algorithm with different values of $\alpha$, when the  average received SNR is  10dB. $\alpha^{[1]}=0.05$, $\alpha^{[2]}=0.2$, $\alpha^{[3]}=0.35$, $\alpha^{[4]}=0.5$, $\alpha^{[5]}=0.65$. From the results, we can see that the average power allocated to each user and parameter $\rho$ of each user can be adjusted according to the values of $\alpha$ set by each user, and thus the expected ID and EH performance can be achieved. The optimal transmitted power of a user becomes larger when its expected rate is larger, and has little relationship with its harvested power. This is because one user can harvest power from all the other users in the network, on the other hand, the rate of a user is almost only determined by its own transmitted power.
Besides, the values of $\alpha$ in the PSO algorithm with power allocation are quite different from those in the PSO algorithm without power allocation according to the ID and EH performance of the users.
\begin{figure}[t]
\centering
\includegraphics[bb=150 292 507 550,scale=.85]{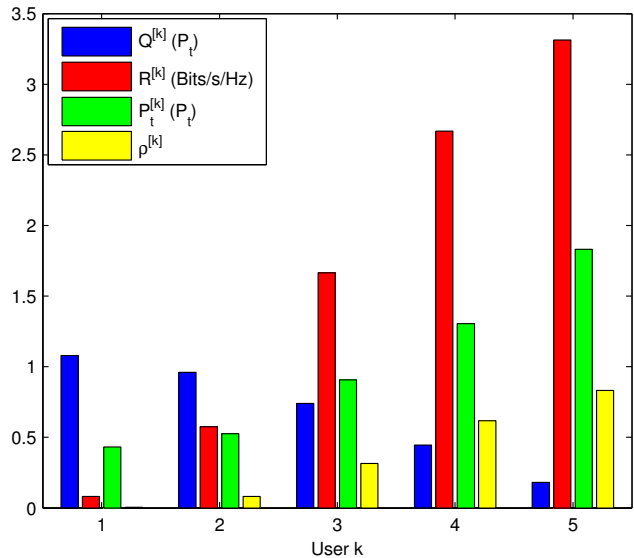}
\caption{The average power harvested, transmission rate, transmitted power allocated, and corresponding parameter $\rho$ of users in the power-allocation PSO algorithm with different values of $\alpha$, when the average received SNR is 10dB. $\alpha^{[1]}=0.05$, $\alpha^{[2]}=0.2$, $\alpha^{[3]}=0.35$, $\alpha^{[4]}=0.5$, $\alpha^{[5]}=0.65$.}
\end{figure}

\section{Conclusions and Future Work}
In this paper, we have presented a common framework of simultaneous wireless information and power transfer in IA networks, and analyzed the performance of SWIPT in IA networks. We derived the lower and upper bounds of the power that can be harvested in IA networks. An easy-implemented SWIPT-US scheme for IA networks was proposed, in which a number of receivers are selected for EH dedicatedly and the others are devoted to ID solely in a time slot. Two algorithms, RRS and PRRS, were designed for SWIPT-US. In the PRRS algorithm,  power-to-rate ratio is defined and adopted to select EH receivers. To continuously optimize the EH and ID performance of each user according to the requirements of the systems when power splitters are available, the PSO algorithm was proposed for IA networks, in which the EH and ID performance is optimized simultaneously over $\rho$. We also studied the  power allocation problem in the PSO algorithm. Simulation results were presented to show the effectiveness and efficiency of the proposed algorithms for SWIPT in IA networks. In our future work, the near-far effect will  be considered in SWIPT for more practical IA networks.

\balance
\bibliographystyle{ieeetr}
\setlength{\baselineskip}{12pt}
\bibliography{ReferenceSCI2}



\ifCLASSOPTIONcaptionsoff
  \newpage
\fi

\end{document}